# REDUCTION AND SIMULTANEOUS DOPING OF GRAPHENE OXIDE TO REPEL LDL IN TREATMENT OF ATHEROSCLEROSIS DISEASE

L.M. Rivera[a,b], A.F. Betancur[b], D.G. Zarate[c], D. Torres Torres[a], L.M. Hoyos[b], A.G. García[a*]

[a] Laboratorio de Síntesis y Modificación de Nanoestructuras y Materiales Bidimensionales. Centro de Investigación en Materiales Avanzados S.C., Parque PIIT, Apodaca Nuevo León, México. C.P. 66628

[b] Grupo de Biología de Sistemas, Universidad Pontificia Bolivariana, Medellín, Colombia

[c] Facultad de Ciencias Biológicas, Universidad Autónoma de Nuevo León, Monterrey, México

* Corresponding author: alejandra.garcia@cimav.edu.mx

**ABSTRACT**

Atherosclerotic disease develops when there is endothelial dysfunction, making low density lipoproteins (LDL) which are cholesterol carriers traveling in the blood, very likely to penetrate the endothelium layer triggering the formation of the atheromatous plaque and leading to the obstruction of blood vessels. For some time now, devices called stents have been used to re-open the light of the plugged artery, allowing the restoration of blood flow. However, this treatment has disadvantages like developing intra-stent restenosis and progression of the atherosclerosis, clogging the blood vessel again. For these reasons it is necessary to develop new materials and strategies focused on the prevention and correction of this pathology and its subsequent complications. In the present work, a reduction and in situ nitrogen doping method in graphene oxide (N-rGO) was studied. By means of XPS a 9.57% at of nitrogen doping was demonstrated and was found that the pyrrole type nitrogen is predominant. Due to nitrogen doping and pyrrolic sites, the N-rGO surface reached high levels of negative electric charge, which was verified by Electrostatic Force Microscopy (EFM) and wettability tests between the LDL at different oxidation levels and the synthesized N-rGO; contact angles between 95.7° and 130.4° were obtained which indicates repulsion between both materials. The N-rGO was also characterized by SEM and TEM where a wrinkled morphology typical of this type of material was observed and the hexagonal atomic arrangement of GO is distinguished. Characterization was also carried out using AFM where thicknesses of up to 2 nm were found, indicating the presence of few layers, in addition by means of FTIR it is demonstrated that the material does not undergo significant structural changes in contact with the LDL. In addition, characterizations by DRX and RAMAN show evidence of the nitrogen presence. Finally, cell viability tests using 3-(4,5-dimethylthiazolyl)-2,5-diphenyltetrazolium bromide (MTT) assay were carried out on human umbilical cord endothelial cells (HUVEC) to analyze the cellular response at different concentrations of N-rGO, obtaining favorable behavior up to concentrations of 100 (μg/mL).

1. Introduction

Coronary artery disease is one of the main causes of morbidity and mortality in men and women, especially in the most developed nations [1]. The rupture of the atheroma plaques generates the formation of clots, which triggers myocardial infarctions, cerebral ischemia and peripheral arterial occlusive disease [2].

The prevalence of diseases such as hypertension, diabetes and even self-imposed conditions such as smoking, among others associated with increased levels of lipids, have promoted the emergence of all kinds of difficulties related to the proper functioning of blood vessels. These pathologies generate an unregulated overproduction of low density lipoproteins (LDL), and increase the risk of endothelial dysfunction.

In consecuence, the progressive deterioration of affected vessels is caused, leading to reduction of blood flow with various complications, additionally symptoms usually manifest in advanced stages of the disease. This pathology involves the diffusion of LDL, which transports cholesterol, into the arterial wall (endothelium), this diffusion is generated due to a disorder called endothelial dysfunction (inability of the endothelium to regulate the production and release of anti-atherogenic vasodilator agents), as a consequence the endothelium increases the production of reactive oxygen species, which oxidize LDL and also increases the expression of LDL modified by oxidation (Ox-LDL) receptors, doing it more permeable to the passage of LDL. Once inside the intima layer (innermost layer of the arterial wall) a complex process is generated by which the arterial wall dilates until partially or totally occludes the artery's light, reducing or impeding the blood flow [3]. Consequently, it is clear that in order to reduce the risk of suffering from cardiovascular diseases, it is necessary to make long-term changes in lifestyle, at the same time new diagnostic methods, monitoring tools and especially new treatments must be developed to help reduce the incidence of said pathologies [4].

Currently the treatment of atherosclerosis is approached from different angles, for example, with the administration of drugs, although due to the irreversibility of the processes generated, no drug has the capacity to reverse the formation of the atheromatous plaque [5]. This pathology has also been widely treated through the implantation via angioplasty of tubular devices with a mesh-like wall, called stents, these devices are positioned in the affected area of the artery and their function is to mechanically reopen the obstructed vessel and restoring blood flow. Some of the stents have been coated with drugs that mitigate endothelial dysfunction or enhance biocompatibility, however it is common to generate the so-called intra-stent restenosis, and the progressison of the atherosclerosis [6] caused by LDL accumulation in endothelial wall. For these reasons, in recent years some studies has been looking for the possible application of different materials that have the facility to modify their surface, among them, carbonaceous materials such as carbon nanotubes in the prevention and treatment of atherosclerotic disease [7]. Nevertheless the high mobility of charge carriers, intrinsic low electrical noise, and reduced cytotoxicity compared to CNTs [8] makes to modified or doped graphene a better candidate for this type of applications.

Therefore graphene based materials offers a good tratment option. Its interesting mechanical, electrical and thermal properties, besides its posibility of modifications and doping [9] makes those kind of materials excellent candidates to reinforce other materials and give them new properties being able to be used in numerous applications.

A great number of methods for a large scale graphene production has been already employed. These methods involve: mechanical cleaving, epitaxial growth, chemical vapor deposition, tearing of carbon nanotubes and reduction of graphene oxide (GO) from graphite and/or exfoliated graphite, direct exfoliation of graphite in a solvent [10], and one of the most used techniques is the chemical oxidation, that produces graphene oxide with a large number of oxygen containing functional groups which attracts many research interest due to its outstanding properties including thermal and chemical stability, high mechanical strength, large specific surface area, good electron conductivity and water solubility [11]. The presence of these oxigen containing groups reduces the thermal stability of the nanomaterial, but may be important to promote interaction and compatibility with other materials such as a particular polymer matrix [12].

Three of the most commun chemical exfoliation methods are: i) Staudenmaier method: fuming nitric acid, concentrated sulfuric acid and $KClO_3$, ii) Hofmann method: concentrated nitric acid, concentrated sulfuric acid and KClO3 and iii) Hummers method: concentrated sulfuric acid in the presence of $NaNO_3$ and $KMnO_4$ [13]. These methods require extensive oxidation of aromatic structure in order to weaken Van der Waals interaction between graphene sheets followed by their exfoliation and dispersión in solution. Resulting GO, multilayered or single sheets have high density OH and COOH groups[14].

Yet one of the most attractive characteristics of GO is that it can be reduced to obtain a material very similar (but not yet equal) to graphene, by removing most of the groups that contain oxygen, leading to the partialy recovery of the original structure. This reduced GO has proved to has more biological compatibility than the GO because it has less oxygen containing groups so it has less possibilities to react with the biological environment. The most common techniques for chemical reduction of GO uses reducing agents such as hydrazine or hydroquinone, sodium borohydride, etc. However, the large scale productions of high purity graphene sheets still remains a challenge [15]. However the graphene sheets prepared via this chemical reduction methods usually retains more defects and uncontrolled geometrical shapes due to the harsh oxidation conditions [14].

The optimization of the exfoliation process using more gentle approaches and other methods are progressing to address this challenging problem toward a different, low waste, scalable and low cost production of GO and graphene [16].

In the present work is reported a new graphene oxide reduction and in situ nitrogen dopping method and its interaction with different oxidized LDL, nitrogen confers to the material a negatively charged surface wich is used to repeal the LDL as show in the wettability analysis by contact angle and EFM studies. This development can be used as a high biocompatible and liporepelent material for stent coatings. Other characterizations as SEM, TEM, AFM, EFM, Raman, FTIR, XPS and DRX were carried out to verify themorphology, layers, functional groups, nitrogen doping and other characteristics of the N-rGO.

2. Methodology and procedures

 2.1 Graphene Oxide Synthesis

GO was prepared by the oxidation of graphite powder. Graphite powder was ultrasonicated with sulfuric and nitric acid (2:1 ratio) in a Branson 1800 bath wich sweeps through a frecuency range of 37 to 43 KHz and a power output of 70 W. Then potassium permanganate was gradually added and heated to 40°C with stirring. After 2.5 hours deionized water was added and temperature was raised up to 100° C. Thereafter deionized water and hydrogen peroxide (9 mL) were also added. Finally the above mixture was left in repose for few minutes and washed with water until a neutral pH was reached.

 2.2 Graphene Oxide Reduction and doping

The reduced and doped nitrogen graphene oxide (N-rGO) was obtained by a process of simultaneus thermal reduction and doping of GO in a 50ml Stainless Teflon Lined Autoclave from Zhengzhou Keda Machinery and Instrument Equipment, which reaches up to 30 bar. A mixture of ammonium hydroxide and neutral pH GO (2.5:1 ratio) was first prepared in the autoclave and heated in a oven to 200° C for 22 hours. After cooling the mixture was washed several times with deionized water until reaches again a neutral pH.

### 2.3 Low Density Lipoprotein Oxidation

Low-density lipoproteins were obtained from porcine plasma with a concentration of 10 μg/ml. The lipoproteins were oxidated at three levels, non oxidized, medium oxidation and high oxidation following the methodology of Evangelia Chnari [17]. The médium and high oxidation were achived by incubation of LDL (50 μg/ml) with 10 μg of copper sulphate ($Cu_2SO_4$) at 37° C during 2 hours and 18 hours respectively. At the end of the oxidation process for both oxidation levels the reaction was completed with an aqueous solution of 0.01% w/v ethylenediaminetetraacetic acid (EDTA).

### 2.4 Cell Viability Assay and cytotoxicity

MTT Assay Kit, dimethyl sulfoxide (DMSO), Dulbecco's Modified Eagle's Medium (DMEM) and, Resazurin were purchased from Sigma Aldrich, all of them used without additional purificacion, Human Umbilical Endothelial Cells (HUVEC) were bought from ATCC, USA. The metabolic activity of human Umbilical Vein Endothelial Cells (HUVEC)s was measured by the reduction of the MTT compound (3 (4,5-dimethyl-2-thiazoyl) -2,5-diphenyltetrazolic bromide), which is taken up by the cells and reduced by mitochondrial succinic dehydrogenase enzyme to its insoluble form formazan. Crystals form inside the metabolically active cells and can be released by the solubilization using dimethyl sulfoxide (DMSO), in this way the amount of MTT reduced by a colorimetric method is quantified, by a change in color from yellow to violet. The ability of cells to reduce MTT is an indicator of the integrity of mitochondria and its functional metabolic activity [18].

HUVEC cells were cultured and maintained in DMEM médium supplemented with 5% FBS and 1% antibiotic / antimicotic in a culture flask and were incubated at 37° C in a 5% CO chamber, Culture medium was changed every 2–3 days removing the previous media. Viable cells were plated in 96-well plates and cultivated until the cell monolayer became confluent. N-rGO powder was dispersed in DMEM by ultrasonication at a concentration of 500 μg/mL. The samples were adjusted to several concentrations: 10, 20 40, 80, 100 μg/mL. A total of 200μIL of the dispersions were added to each well and cultured for either another 24 h, 48 h and 72 h at 37° C. Untreated cells were used as a positive control and cells treated with 1% Triton X-100, a toxic compound which does not affect MTT reduction as a negative control. Next, 100 µL of MTT were added to each well and the samples were incubated for another 2 h at 37° C, then 20 µL of DMSO were added to each well and the absorbance of the solution at 570 nm was recorded with a Biotek Synergy HT microplate reader. Every experiment was performed in sextuplicate.

Resazurin is a redox dye commonly used as an indicator of chemical cytotoxicity in cultured cells normally is a blue non fluorescent dye that turns into resorufin (red and fluorescent) when reduced, it is and indicator of celular growth, because cell proliferation creates a reduced enviroment while inibition of proliferation creates an oxidated enviroment. The assay is based on the ability of viable, metabolically active cells to reduce resazurin to resorufin [19]. For Resazurin assay HUVEC cells were cultivated as explained before, same N-rGO concentrations were used, untreated cells were used as a positive control and anti-T cell immunotoxins (IT) as a negative control [20]. After 24, 48 and 72 h, 10 µL of resazurin were added to each well and the fluorescence of the solution at 530 nm was measured with a Biotek Synergy HT microplate reader. Every experiment was performed in sextuplicate. The percentage of cell viability for both MTT and Resazurin was calculated using the following formula:

$$\text{Cell Viability (\%)} = \frac{\text{Number of Viable Cells}}{\text{Number of live + dead cells}} \times 100$$

## 3. Characterization

Scanning Electron Microscope (SEM): a Fei Nova NanoSEM 200 scanning electron microscope was used. The microscope was operated at 7 kV, and images were taken at 2.000, 5.000 and 100.000 X. Scanning Transmission Microscope (TEM): a JEOL JEM 2200FS+CS TEM was used. The microscope was operated at 200 kV. Raman: measurements were taken in a LabRam HR Evolution from Horiba at ambient conditionswith controlled light and a back-scattering configuration at a wavelength of 532 nm. Atomic Force Micoscopy (AFM): a MFP3D-SA microscope from ASYLUM RESEARCH in non contact mode was used, scannings were done in areas of 80 x 80 µm, with velocity range from 0.20 to 0.75 Hz. A silicon cantilever with a rectangular geometry of 240 mm of length, completely coated with a Ti/Ir film model AC160TS-R3 from ASYLUM RESEARCH was used with first nominal resonance frecuency of 300 kHz and nominal spring constant of 26 N/m.

Electric Force Micrscopy (EFM): a MFP3D-SA mifrom ASYLUM RESEARCH in non contact mode was used, In order to have an accurate detection of electric responses between the cantilever and sample surface, its separation distance should goes at least through van der Waals (VdW) forces magnitudes. In this regards, it is well know that VdW forces are generally detectable at a smaller cantilever-sample surface distance than 20 nm [21]. Therefore, in the present work a distance up to 50 nm allowed $z$-component electric force gradient suitably detectable, when a voltage of $Vs$ = 3 volts was applied between tip and sample surface. X-Ray-Diffraction (XRD): a Panalytical model Empyrean X Ray Difractometer with a K-Alpha1 wavelength of 1.540598 Å, a scan step size of 0.017, a time per step of 59.69 and a Generator voltaje of 45 V was used. X-ray photoelectron spectroscopy (XPS): a ThermoScientific model Escalab 250Xi photoelectron espectrometer was used with excitation source of monochrome aluminum k-α, vacuum, pass energy of 20 eV and load compensation cannon, the "take-off" angle was 45°. Goniometer: a OCA 15 Plus from dataphysics was used to measure contact angle between pastilles made of both doped and rGO and GO by using a mold in a mechanical press and LDL with different levels of oxidation. FTIR: a iS50 from Thermo Scientific infrared spectrometer was used.

## 4. Results and Discussion

### *4.1 Nitrogen Doping Reaction Mechanism*

After GO synthesis a simultaneous process of reduction and in situ nitrogen doping was accomplished. The reaction mechanism proposed for this doping is described below. As ammonium hydroxide $NH_4OH$ is ammonia combined with water, the reaction chamber contains these compounds ($NH_3$ and $H_2O$). The autoclave reaches a maximum internal pressure of 30 bar and a temperature of 200° C, therefore the water present in the reaction is in liquid state according to its phase diagram [22], indicating that water may not be actively participating in the doping reactions and is just the medium where the reaction takes place.

On the other hand, the decomposition of ammonia subjected to thermal treatments is complicated, it is known that hydrogen and nitrogen are produced, which after cooling and through various reactions can generate as resulting products $NH_2$, NH and N as shown in Figure 1 a. Since the decomposition temperature of $NH_3$ at atmospheric pressure is 500° C is essential to use an autoclave to increase the pressure and thus achieve decomposition at a lower temperature that does not negatively affect the GO.

The decomposition of $NH_3$ is not only dependent on the temperature, but also of the nature of the contact surface. White and coworkers experiments prove that the greater the surface contact, the decomposition of $NH_3$ is enhanced, since the process is only catalyzed when the molecules come in contact with a surface [23]. We suggest that this happens since all surfaces must be passivated, normally with terminal OHs, when $NH_3$ reacts with OH it begins to oxidize, and new reactions are catalyzed that trigger the decomposition process, hence, a big surface area means greater amount of OHs and greater degree of decomposition; this can apply for the presence of oxygen and hydrogen containing functional groups, such as those present in the GO, as the XPS results show. Therefore, graphene oxide, which has a very high surface area, influencing the degree of decomposition of $NH_3$, allowing the generation of more hydrogen and nitrogen available for recombination and subsequent doping, this in conjunction with the temperature and pressure of the process can explain the high percentage of nitrogen bound to carbon.

Once formed the products obtained by the recombination of nitrogen and hydrogen ($NH_2$, NH and atomic N), several events as explained below could describe the generation of the bonds between the carbon network and nitrogen.

It is well known that the native structural defects, nanoholes and vacancies in the graphene sheets have a relatively high activity; among the defects and vacancies most commonly generated are the (55-77) commonly called Stone-Wales, the vacancy (5-9) and the divacancies (5-8-5) and (555-777) [24].

The structure of the defects in combination with the kinetic and thermodynamic processes participates in the complex reorganization reaction that is carried out for the generation of quaternary nitrogen, pyridine and pyrrolic nitrogen.

Recent studies using ab-initio molecular dynamics simulation have shown that some vacancies have the appropriate electronic structure to trap nitrogen-containing groups and mediate a subsequent dehydrogenation process. The reason why vacancies can trap such groups to nitrogen atoms is the presence of 2 types of localized force fields, the first known as the field of tension is caused by the imperfections and disorder that the vacancy generates in the network leaving it in a state of non-equilibrium, willing to find new chemical bonds to stabilize itself. The second is called electronic field and is generated by the existence of loose bonds, this field only have the vacancies generated by the loss of a carbon atom [25].

All vacancies or defects have different levels of localized force fields depending on their formation characteristics and their final structure, for example the native defect Stone Wales (55-77), which is formed by the rotation of 90° of a C-C bond only has a voltage field and itis not strong enough to trap the nitrogen atoms or groups. On the other hand, the vacancy (5-9) is formed by the loss of a carbon atom with three loose bonds, so in this vacancy both the field of tension and the electronic field are generated reaching the strength to attract groups with nitrogen and atomic nitrogen.

The divacancy (5-8-5), which although formed by the loss of 2 carbon atoms leaves no loose bonds because its structure favors a process of self-saturation [26], thus, it only has the stress field and fails to attract nitrogen groups or individual atoms. Another example is the divacancy (555-777) which can be formed by the coalescence of two simple vacancies (5-9) or also by the 90° rotation of a C-C bond of a divacancy (5-8-5) as this defect can rotate and migrate in the graphene plane through a sequence of Stone-Wales type transformations [27]. This divacancy lose 2 carbon atoms and has no loose bonds, however this rotation decreases the (5-8-5) vacancy nanohole but also greatly increases the disorder in that region of the network, meaning that the field of tension is more powerful than the one of the divacancy (5-8-5) and even without electronic field can attract the groups and atoms of nitrogen. The divacancy (555-777) has the less formation energy respect to the mentioned defects, for this reason it is expected that its presence will be dominant in the network, therefore good possibilities of forming bonds between the carbon and the nitrogen can be expected.

$NH_2$ is attracted by the force field of a divacancy (555 -777) and remains on its surface until another $NH_2$ is also attracted and interacts with the first (Figure 1 b), this interaction forms NH and $NH_3$ (Figure 1 c), as $NH_3$ cannot be trapped by the defects it goes to the surface and remains drifting (Figure 1 d); on the other hand the NH stays in the vacancy until another NH approaches (Figure 1 e), from this second interaction N atomic and NH2 are obtained (Figure 1 f), the N atomic combines with the first pair (5-7) introducing in the middle of a C-C bond and re-forming a hexagon, (N pyridinic) as shown in  Figure 1 g, finally represented in Figure 1 h, the $NH_2$ moves to the second and third pair (5-7) that absorbs it in a cyclic process [25].

Later the formation of quaternary nitrogen can be achieved from the pyridinic N, this reaction is only possible form the third and last nitrogen anchored to the third (5-7) pair, this nitrogen is known as N-bridge, since this is too unstable it can react easily with another N-bridge nearby or with an atomic N that gets close enough to form the nitrogen molecule ($N_2$) with which it desorbs from the surface; this reaction catalyzes a series of rotations, breakdowns and formation of new bonds that find their equilibrium state forming 2 quaternary nitrogen [25].

Regarding the formation of pyrrole N, it can be presumed that is generated by a similar mechanism, taking advantage of the armchair configurations of the edges of the sheets with four carbon atoms, the NH is positioned at the edge passivating the loose bonds present in these edges and forming a pentagon.

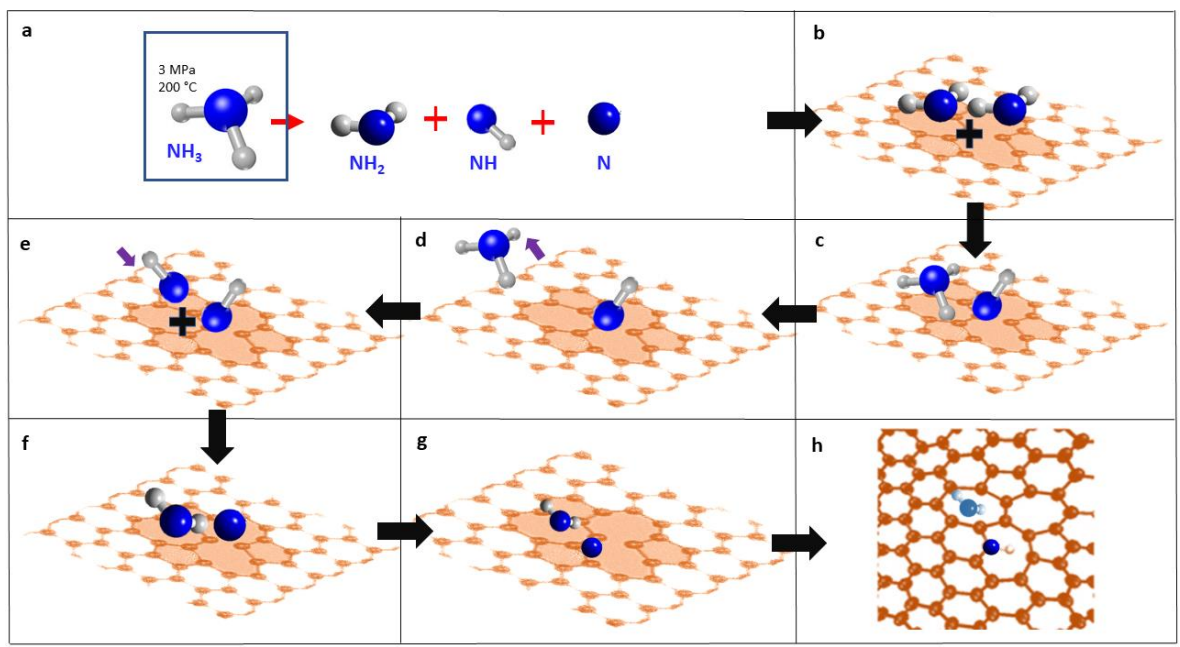

*Figure 1 Nitrogen doping process in rGO lattice.*

### 4.2 Scanning electron microscopy (SEM)

SEM images (Figure 2a and Figure 2b) show the prescence of surface corrugations such as wrinkles and ripples for both GO and N-rGO, evidencing that there is no significant structural change that negatively affects the GO once reduced and doped. This corrugations are tipic in 2D nanomaterials with lattice defects or surface anchored functional groups that modify the $sp^2$ hybridization, they can be formed by thermal vibrations, edge inestabilities, themodinamically unstable interatomic interaction or thermal contractions [28]. There is also is a thermodinamic requirement for the existence of out-of-plane bendindg in crystals [29], therefore 2D crystals present some kind of inestability that is compensated by ripple and wrinkles formation induced by changes in the C-C bond lenghts due to interatomic interactions and thermal vibrations that induce carbon to occupy space in a third dimension [30], this minimize the total free energy generating the formation of such surface corrugations [31]. There is also possible to notice a difference between the GO and N-rGO surface, the first one has wider and bigger wrinkles while the second one presents thiner and more intricate wrinkles, this behavior is possible due to the loss of most of the fuctional groups during the reduction process and the introduction of the different type of bonds with nitrogen in the lattice.

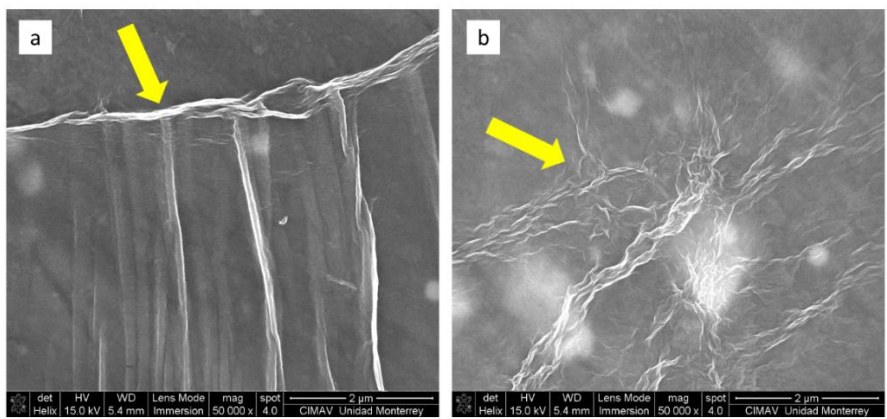

Figure 2. SEM images of a) GO, b) N-rGO.

### 4.3 Transmission Electron Microscopy (TEM)

High resolution transmission electron microscopy of GO (Figure 3 a) and N-rGO (Figure 3 b) show again, corroborating SEM results, the presence of ripples and wrinkles in the sheets, the insert in Figure 4 a shows a STEM image of a GO flake, the left insert shows the fast Fourier Transfom (FFT) of the area enclosed in yellow, wich exhibits a definied hexagonal pattern indicating the existence of one graphene layer, the right insert shows a closer view of the area enclosed in red, the hexagonal pattern and C atoms can be apreciated, there is also presence of some imperfection in the lattice as well as amorphous zones due to the exfoliation process (appreciated by amorphous diffraction pattern). Figure 4 b presents a STEM image of N-rGO and the insert correspond to the fourier transorm of the entire image where the hexagonal pattern can be also seen. N-rGO display distinct Moire´ patterns caused by the stacking of individual hexagonal layers with different orientations, then the occurrence of these hexagonal patterns implies a long-range orientational hexagonal order in the sheets [32]. The FFT of this sample exhibits a better diffraction pattern compared with GO correlating with oxygen groups lost.

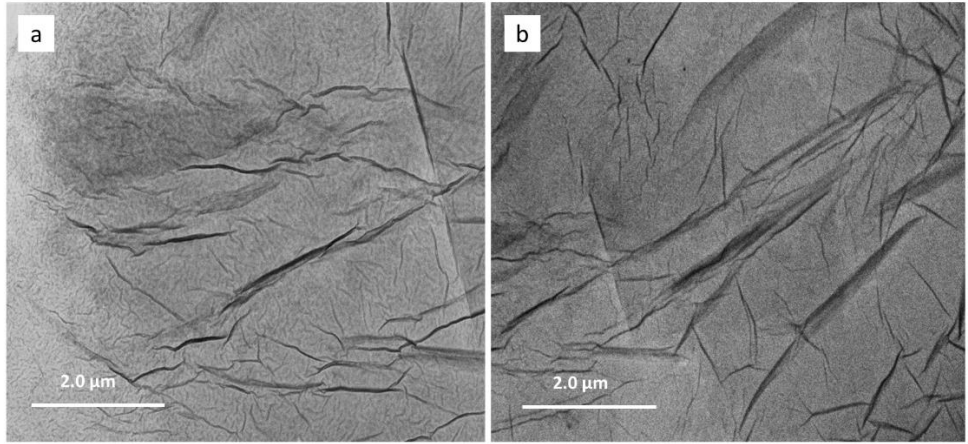

*Figure 3. TEM images of a) GO, b) N-rGO.*

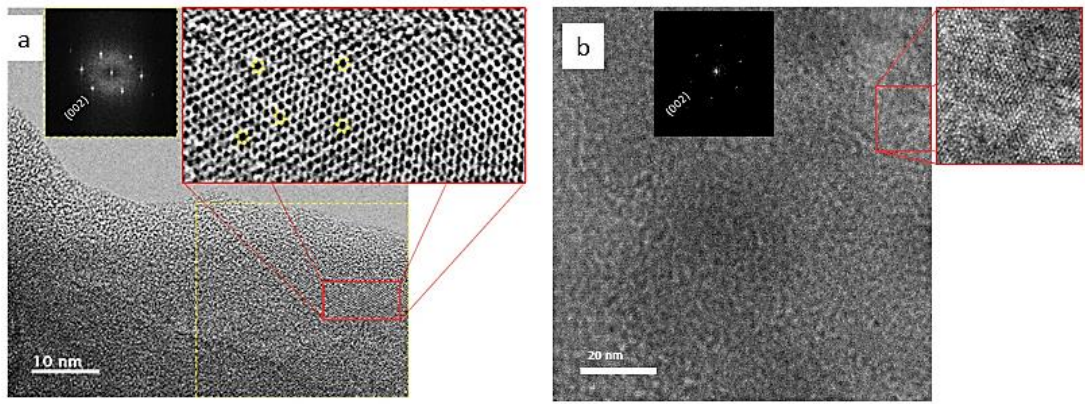

*Figure 4. TEM images of a) GO; right insert corresponds to the red area zoom, left insert corresponds to the Fourier transform of the yellow area, b) N-rGO; right insert corresponds to the red area zoom, left insert corresponds to the Fourier transform of the red square.*

### 4.4 Raman Scattering

Raman spectra of GO and N-rGO using 532 nm laser excitation are shown in Figure 5. In both samples bands D, G and 2D can be distinguished; the D band indicates the presence of defects in the graphitic structure [33], the G band corresponds to the first order bond stretching of $sp^2$ carbon atoms ($E_{2g}$ mode) and the 2D peak is attributed to double resonance transitions resulting in production of two phonons with opposite momentum. This band is different from the D peak, which is only Raman active in the presence of defects, the 2D mode is active even in the absence of any defects [34]. The presence of 2D gives and indicative of the number of layers in graphene [35], when the peak is very sharp and intense means one or very few stacked graphene layers, for GO and N-rGO 2D band is widened meaning that there are more stacked layers. It is reported that the $I_{2D} / I_G$ ratio can give an estimative of the number of layers; a $I_{2D} / I_G > 2$ gives an indication of one layer, if $1 < I_{2D} / I_G > 2$ it could be a bilayer, finally if $I_{2D} / I_G < 1$ indicates a possible multilayer material [36]. The GO and N-rGO have a $I_{2D} / I_G$ ratio of 0.69 and 0.33 respectively, thus acording to the last statement both samples are possibly few layers graphene.

D band is at 1367 cm$^{-1}$, and 1350 cm$^{-1}$ for GO and N-rGO respectively, D band is at 1602 and 1599 cm$^{-1}$ for GO and N-rGO respectively. Compared to that of GO, the Raman spectrum of N-rGO shows a downshift of the G-band (3 cm$^{-1}$) and 2D band (6 cm$^{-1}$). Based on the empirical relationship between the Fermi energy and the Raman peak position, the downshifts suggest that the N dopants move the Fermi level of graphene [37].

Also the D-band and G-band peak intensities ratio ($I_D/I_G$) of Raman spectra provides information about the structural changes and defects since an increase in any disorder in the graphene will cause $I_D/I_G$ ratio to increase [38]. This ($I_D/I_G$) ratio is 0.92 for GO and 1.14 for N-rGO; this means that N-rGO has higher degree of deffects than GO, due to the modification that nitrogen causes in the lattice. Nitrogen doping can also be related to $I_D/I_G$ rate, an increase in this value can be explained by the elastically scattered photo-excited electron created by the large number of nitrogen atoms embedded in the graphene lattice before emitting a phonon [37].

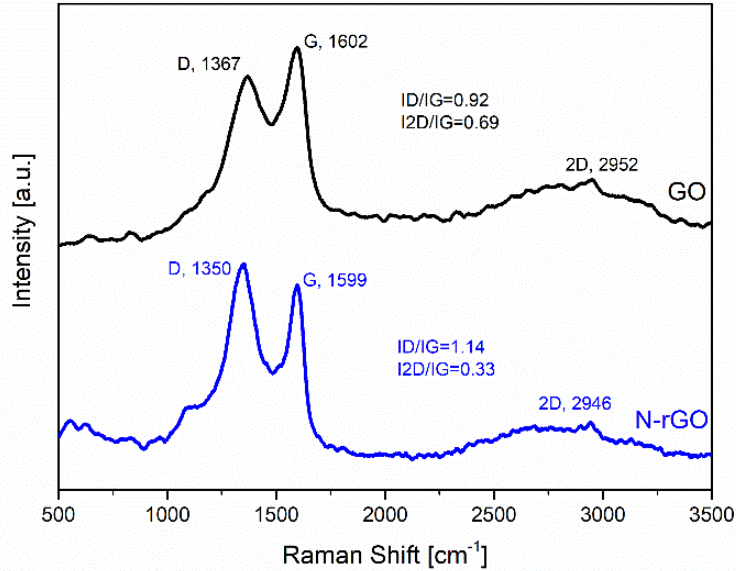

*Figure 5. Raman Spectra of GO and N-rGO, 532 laser.*

### 4.5 Atomic Force Microscopy

Atomic force microscopy (AFM) was used to evaluate the morphologies of the GO and N-rGO, Figure 6 shows typical AFM images of the flakes and Figure 7 shows their corresponding transversal profiles. Images were procesed with WSxM 4.0 software [39].

It should be noted that there are flakes with an average thickness of 2.5 nm as shown in the transversal profiles of Figure 7a and Figure 7b for GO and Figure 7c and Figure 7d for N-rGO, indicating that material was almost fully exfoliated into individual sheets, taking into account that the atomic radius of carbon is 0.15 Å and the thicknes of the oxigen functional groups is in average 0.930 nm [40], the presence of 2 to 3 layers off graphene stacked can be calculated. The images also show that there are some flakes with areas in the order of the microns and there are also smaller flakes with areas in the order of the nanometers on top of the bigger ones (line 4 in Figure 5 b).

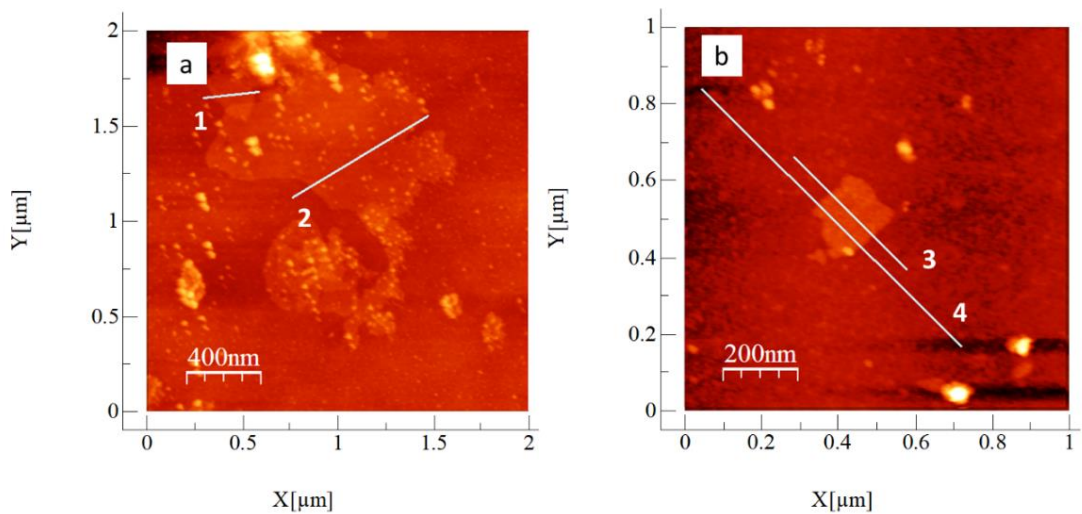

Figure 6. AFM of a) GO, b) N-rGO.

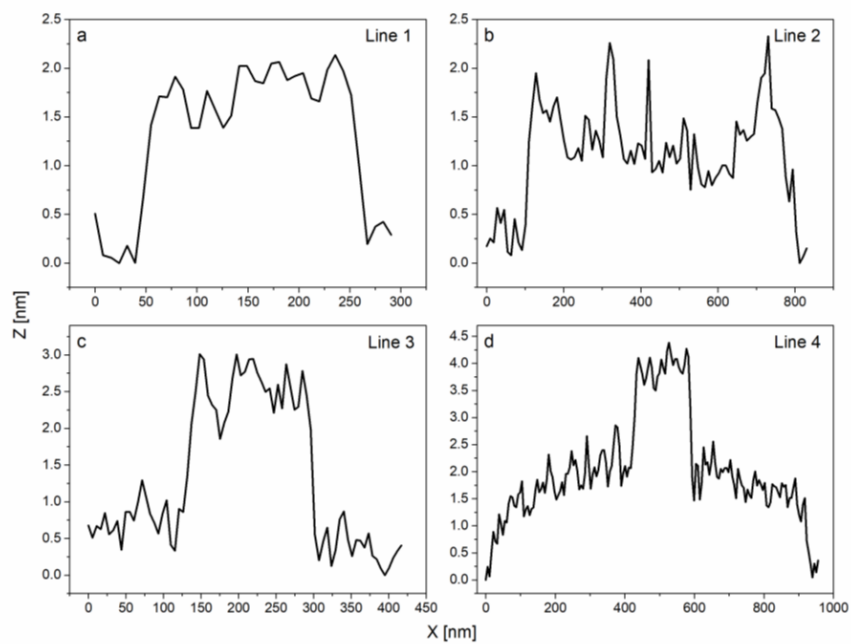

Figure 7. Transversal profiles from AFM of a) GO line 1, b) GO line 2, c) N-rGO line 3, d) N-rGO line.

*4.6 Electric Force Microscopy (EFM)*

The typical surface topography of the N-rGO on Figure 8 a was obtained in AC-mode scanning. A topographical edge consisting of staked N-rGO sheets, leads to the identification of the location of the analyzed sample surface area marked between the two blue points. The corresponding sample surface profiles of height related to the sheets edge topography can be observed from Figure 8 b as red line. Clearly, a well-defined step of about 20 nm can be observed.

Afterward, with the aim to analyze the electric response exhibited by the N-rGO sheets, the EFM scanning in AC-mode was carried out, see Figure 8 c. In order to get the cantilever-sample surface interaction with a high sensibility, cantilever deflections were also analyzed by means of phase shift grades quantification. Firstly, the morphology concern to phases shift from topography relief was achieved. Thus, the corresponding sample surface profiles of phases shift related to the N-rGO sheets edge can be observed from Figure 8 d , where the regarding phase shift of about 9° can be observed in a suddenly form. However, once the contact with the sheets step is overcome, the phase shifts has an immediate reversibility.

Right away, the analysis of cantilever phases during the EFM scanning in AC-mode was carried out at the same sheets surfaces location, see Figure 8 e.  In the present case, changes from bright to dark color concerns to changes in phase of the cantilever vibration. This reveals that the tip perceives a force gradient over the substrate enclosed by each GO sheet, and clearly in a different way between the one below and the one above, as yellow and purple colors, respectively.

Accordingly, the cantilever interaction with the above-sheet (0.4°) results in a tangible torsional force, due to the electric field is acting in an "out-plane" signal. In the other hand, the cantilever interaction with the below-sheet (-0.4°) results in a tangible torsional force but acting in an "in-plane" manner. This force cannot be attributed to a mere topographic effect since in EFM scanning, values of the cantilever amplitudes or deflections always refer to the condition "in air". Hence, in the present research the cantilever lifted was about 50 nm above the sample surface. As a result, in the respective EFM scanning, the additional values of changes in amplitude or deflection signals are subject to deviations, but only due to the attractions or repulsions forces of electric character between the cantilever and sample surface. The corresponding phases shift profiles when analyzing its opposing directions are shown in Figure 8 f.

Consistently, maximal cantilever-phase shifts appear not to have a reversibility behavior once the contact with the sheets step is overcome as in the topographic form, but on the contrary, the phase change remains stable at the new phase shifts magnitude according to the nature of charge trapping by the N-rGO sheets.

Regarding to the calculation of z-component electric force, a value of about $\partial F_z / \partial z$ = 2.16 N/m was obtained. Here were used a spring constant k = 2.6 N/m, a quality factor about Q = 172 from the cantilever calibration, as well as an electric phase shift of $\varphi$ = 0.4° measured at the edge between flat areas with well-defined sheets boundaries due to electric differences from EFM micrograph on Figure 8 f. Therefore, it was demonstrated the exciting new capabilities of the full-featured spatial distribution of N-rGO sheets with remarkable electric feature due to the bimodal nature of N-rGO distribution obtained, characteristics that improves the lipoprotein repulsion as a function of its negative electric charge.

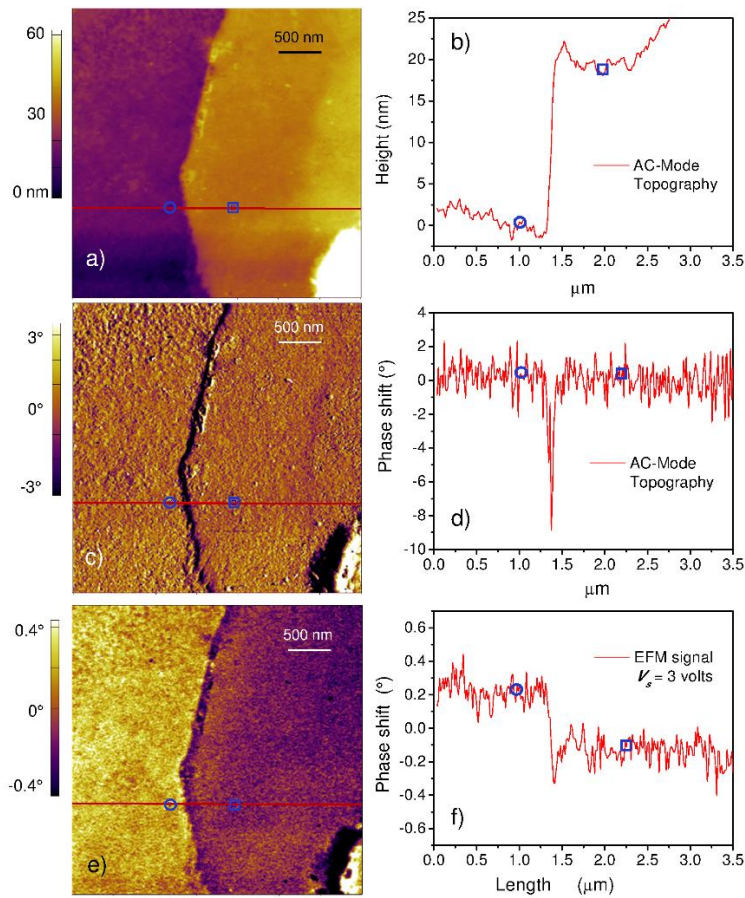

*Figure 8. a) N-rGO surface topography, b) N-rGO surface profile of height, c) EFM scanning in AC-mode, d N-rGO sheets edge surface profiles of phases shift), e) EFM phase image, f) EFM phases shift profile.*

### 4.7 X Ray Diffraction (XRD)

The XRD spectra measured in a range of $2\theta$ from 5° to 70° (Figure 9) shows a comparison between graphite, GO and N-rGO X ray diffraction pattern. For graphite a sharp diffraction peak at $2\theta$ =26.6 can be seen and also a small peak at $2\theta$ =44.5°, typic contributions in graphite, the first one corresponds to the diffraction of the (002) plane with an interplane distance of 3.43 Å; for GO this (002) peak deacrease in intensity and a new diffraction peak at $2\theta$ = 10.6° correspondinig to the (001) plane appears, the interplane distance for GO after exfoliation is 8.37Å , finally for N-rGO the first diffraction peak shifts to $2\theta$ =26° and the second one remains at $2\theta$ =42.3, indicating a short range order in stacked graphene layers [10]. Nevertheless is worth noting that the contribution at $2\theta$ =26.6° is wider tan the one at the same position for the graphite, this could be attributed to lacks stacking secuences compared with graphite before oxidation and reduction process, the interplanar distance for N-rGO is 3.51 Å originated from the (002) plane. All interplane distances were calculated from the Braggs law, with a wavelenght of the x-ray of 1.54 Å,

$$n\lambda = 2d\sin\theta$$

Were n is an integer representing the order of the diffraction peak, $\lambda$ is the wavelenght of the x-ray, *d* is the interplane distance and $\theta$ is the scattering angle.

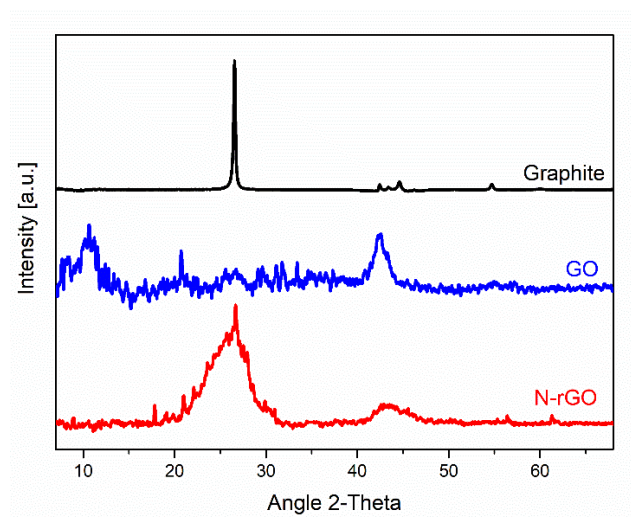

Figure 9. XRD of graphite, GO and N-rGO.

### 4.8 X Ray Photoelectron Spectroscopy (XPS)

By means of the procedure implemented in this work, a doping of 9.57 % at of nitrogen was achieved in the N-r GO sample, in order to achieve a notorius change in the electric surface charge for the specific biological application a doping percentage above 5% (that is the highest achived by most of the published thermal GO reductions) is desirable [44].

Several parameters were taken into account for the deconvolution of curves to ensure greater certainty of the reliability of the fits; data were processed using the software AAnalyzer, that works with the active method, which does not generate user dependency due to its ability to adjust all parameters simultaneously. In addition, both, the type and the combination of backgrounds employed were taken into account, commonly for the deconvolution of the C1s the shirley sherwood type background was used, however it should be considered that the background is associated with energy losses of the electrons and these can be both intrinsic (events within the atom) as extrinsic (events in the crystalline structure), therefore is necessary to have a combination of these backgrounds representing both types of losses, for such reasons the present deconvolutions uses the shirley sherwood type background, which has an intrinsic origin, and the slope background with extrinsic origin, the slope is a more current background that replaces the touggard, that requires to have at least 30 ev distance after the peaks, sometimes making adjustment difficult.

The type of curve used for each contribution is also taken into account, among others the Gaussian, Lorentzian, Voigt and Double Lorentzian curves can be options, the first one corresponds to extrinsic effects related with the equipment, the second one corresponds to intrinsic characteristics of the sample, for such reasons the voigt curve is usually used, which is the Gaussian and Lorentzian convolution, however it is very important to keep in mind that the voigt curve is not well adjusted to all contributions since it is suitable for symmetric peaks and if it is used in asymmetric peaks could create the illusion that the peak can be deconvoluted into two or more peaks, normally for asymmetric peaks doniach-sunjic type curves were used however, these are not integrable so the double lorentzians is used. As the C sp$^2$ contribution is asymetric, the double lorentzian curve was used, for all the other contributions the voigt curve was used ensuring the contribution of both, gaussians and lorentzians curves by means of the stablisment of restrictions. For the reliability of the results elements bonding are verified not only in the C1s but also in the O1s and the N1s [41]. Finally, two main indicators of the quality of the fits are the graph of residuals wich indicates if there are possible missing and the chi square that is a statistical indicator that relates the difference between the experimental point and the model, the lower the Chi Square the better the fit.

High resolution scan and deconvoluted peaks in the C 1s region for GO is shown in Figure 10 a, after reduction/doping process (N-rGO) the C-C sp$^2$ peak intensity increases due to the restauration of the lattice (37% at C for GO and 75.77% at C for N-rGO) and the intensities corresponding to the bonds between C and oxygen functional groups decreases due to the succesfull reduction procces (63% at O for GO and 15.32% at O for N-rGO) Figure 10 c. N-rGO also presents the contribuitions of the C=N sp$^2$ bond at 285.3 eV and C-N sp$^3$ at 287 eV, it is important to notice that unlike N-rGO the GO does not present bonds with nitrogen, corroborating that the nitrogen introduction is atributed to the reduction/doping process. The mentioned peaks corresponding to the bonds between C and O are corroborated for both samples with the high resolution scan in the O1s region, shown for GO in Figure 10 b, where the C-O, C-O-C and the O-C=O functional groups are also present and Figure 10 d for N-rGO where the same functional groups are present (remanents of the reduction process) and also a contribution for the bond between C and N can be seen.

Finally the high resolution scan in the N1s region for the N-rGO sample is presented in Figure 10 e. As well as chemisorbed N at 405.6 eV and oxidized N at 402.7 eV, presence of three types of N-C bonds can be identified, pyridinic N sp$^2$ hybridized at 398.1 eV, pyrrole N sp$^3$ hybridized at 399.5 eV, with the biggest area, and quaternary or graphitic N sp$^2$ hybridized at 401.3 eV, this bonds contribute with 1, 2 and 1 electron to the $\pi$ cloud respectively. The presence of this C-N bonds, specially de pyrrolic N wich is the one in greatest quantity, are the reason why the entire laticce surface acquieres a very negative electric character, necesary for the repulsion of the mentioned LDL wich also have negative electric character.

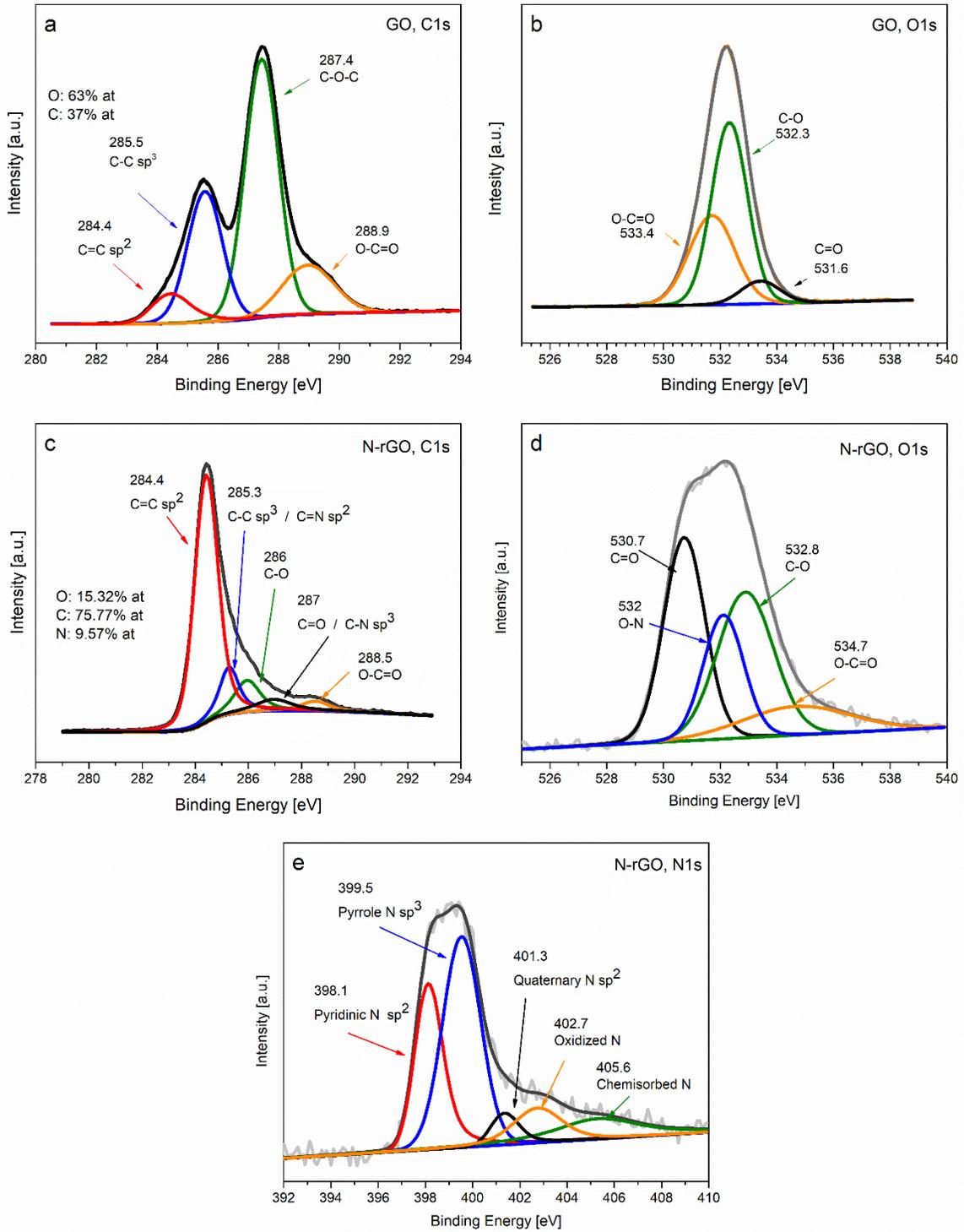

*Figure 10. XPS of: a) C1s for GO, b) O1s for GO, c) C1s for N-rGO, d) C1s for N-rGO, e) N1s for N-rGO.*

### *4.9 Contact Angle Measurements*

Oxidized LDL, commonly known as oxLDL, results from the exposure of LDL to a number of oxidizing agents such as superoxide anions or hydrogen peroxide present in cells, particularly macrophage cells of the arterial wall, enzymes such as lipoxygenases and myeloperoxidase products; this exposure can lead to the depletion of antioxidant compounds and subsequent oxidation of the lipids and proteins present in the LDL [42]. In addition, the oxidation of lipoproteins promotes the expression of proinflammatory genes that cause the dysfunction of vascular endothelial cells [7], so it is important to find mechanisms to keep said oxidized lipoproteins away from the wall of blood vessels, in order to contribute to the improvement of endothelial dysfunction.

The contact angle test was carried out to verify the differences in the repulsive behavior of N-rGO towards LDL with different oxidation states, simulating in this way different levels of atherosclerotic disease progression. Figure 11 a shows that the contact angle between GO and LDL is 18.9°, meaning that the material is lipophilic, therefore GO without nitrogen content cannot repel the negative charges of LDL. On the other hand from Figure 11 b, a 95.7° contact angle between non-oxidized LDL and N-rGO can be observed meaning that N-rGO is lipophobic, it can repel negative charges. For the LDL with medium oxidation degree and N-rGO (Figure 11 c) the contact angle was 121.7° still in the range of lipophobic behavior but clearly generates more repulsion; and finally, the highly oxidized LDL generates the higher contact angle with the N-rGO (130.4°) as seen in Figure 11d, which is also lipophobic. These results show a clear increase in the repulsive behavior of N-rGO with the increase of the LDL oxidation degree, hence there is an indication that the final material will have an important effect in all stages of atherosclerotic disease.

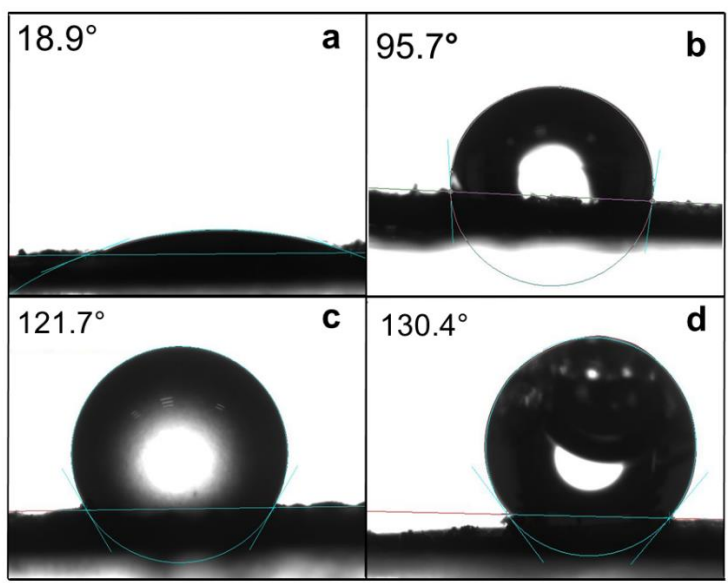

*Figure 11. Contact Angle between a) GO and no oxidized LDL, b) N-rGO and non-oxidized LDL, c) N-rGO and medium oxidized LDL, c) N-rGO and high oxidized LDL.*

### *4.10 FTIR*

Characterization by FTIR in ATR mode was carried out for the N-rGO and N-rGO moistened with LDL. Figure 12, shows that GO has the characteristic belly form of the OH at 3300 cm$^{-1}$, this contribution is also observed in the N-rGO and the N-rGO with LDL, in addition GO presents another contribution at 1734 cm$^{-1}$ which is attributed to the stretching vibrations of the C=O belonging to the COOH groups and another one in 1620 cm$^{-1}$ that can be attributed to the presence of non-oxidized graphitic domains or adsorbed water molecules [43]. It can also be seen for GO a peak formed by several contributions, among them at 1224 cm$^{-1}$ corresponding to the stretching vibrations of the C-OH [44]. Subsequently the contribution of 1050 cm$^{-1}$, corresponding to stretching vibration of the C-O in the C-O-C [44] and finally for the GO the 938 cm$^{-1}$ contribution corresponding to the wagging of the OH outside the plane is observed.

For N-rGO the most characteristic peak is observed at 1460 cm$^{-1}$, which corresponds to the C-N sp$^3$, this type of bond is responsible for the wrinkles presented in SEM images [45]. There is also a widened peak that contains several contributions, the C-O at 930 cm$^{-1}$ is slightly displaced from the position in which it was in the GO, this displacement could be attributed to the action of the dopant, the widening of said peak is due to the overlap of NH$_2$ signal appearing at 760 cm$^{-1}$. The contributions related to nitrogen, which appears for N-rGO but not for GO, are an indicative of the presence of the doping element in the material. Finally, for N-rGO with LDL a similar behavior to N-rGO was observed, with the same contributions, however it is noticeable that the intensities of the peaks are lower, this is attributed in part to the action of the LDL, specifically the presence of the plasma amides that appear at 1670 and 1600 cm$^{-1}$ [46]. A very particular characteristic of N-rGO and N-rGO with LDL is the presence of two peaks at 1598 cm$^{-1}$ and 1241 cm$^{-1}$ for N-rGO and 1640 cm$^{-1}$ and 1234 cm$^{-1}$ for N-rGO with LDL.

FTIR characterization demonstrates that the contact of LDL with N-rGO does not represent a risk of modification of its characteristics and properties.

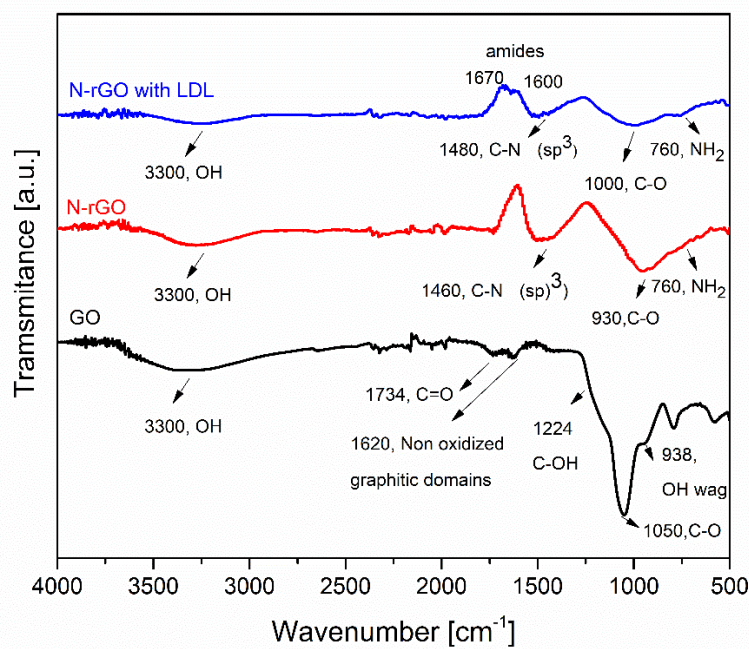

*Figure 12. FTIR of GO, N-rGO and N-rGO moistened with LDL*

### 4.11 Cytotoxicity test

MTT assay and Resazurin assay for HUVEC cells were performed with six different N-rGO concentrations, 10, 20, 40, 60, 80 and 100 µg/mL, and 24, 48 and 72 hours of treatment. The metabolic activity of HUVEC cells measured by MTT assay is shown in Figure 13a and viability behavior related with cell growth with N-rGO treatment measured by Resazurin assay is shown in Figure 13b. It is possible to observe a decreasing tendency in the percentage of viability with the increase in concentration and time, this behavior is maintained in all treatment periods studied. The N-rGO has a strong tendency to agglomeration and this tendency increases at higher concentrations; this behavior interfere with the interpretation of the MTT and Resazurin tests for concentration above 100 µg/mL, for that reason, an extrapolation of the experimental data was carried out in order to know the concentration of N-rGO where the cells response is reduced by half (IC50); According to the extrapolated data (Figure 14 a) HUVEC cells reaches IC50 at 213, 186.1 and 207.2 µg/mL for 24, 48 and 72 hours of treatment with MTT assay. It is notorious that samples treated for 48 hours reached the IC50 first than samples treated for 72 hours, this behavior can be explained by the agglomeration tendency of N-rGO, it is possible that at higher concentrations and long treatment time the material was deposited only on a part of the culture well, leaving some cells without contact and therefore without the influence of N-rGO.

It should be noted that the implementation of the resazurin method is crucial to assess in detail, not only the cytotoxic effect of products, but the behavior of the cells after of being subjected to a treatment and allows to complement MTT results for the evaluation of cell viability by means of determination of the proliferative capacity of cells after specific treatment [47].

From Resazurin assay results in HUVEC cells, extrapolated IC50 of 386.2, 329.5 and 313.4 µg/mL (Figure 14 b) for 24, 48 and 72 hours of treatment were obtained, these concentrations are higher than the obtained for MTT assay, but not significantly different indicating that there is a slightly tendency of N-rGO to generate more significant changes in cell proliferation. Nevertheless, after the 24 hours treatment period, and concentrations until 100 µg/mL the results indicate that the compounds did not cause a significant decrease in cell viability, nor using the MTT method or the resazurin method giving an indication that the N-rGO will not adversely affect endothelial wall cells.

Further study with other cell lines is needed to determinate the influence of the interaction of N-rGO with blood cells.

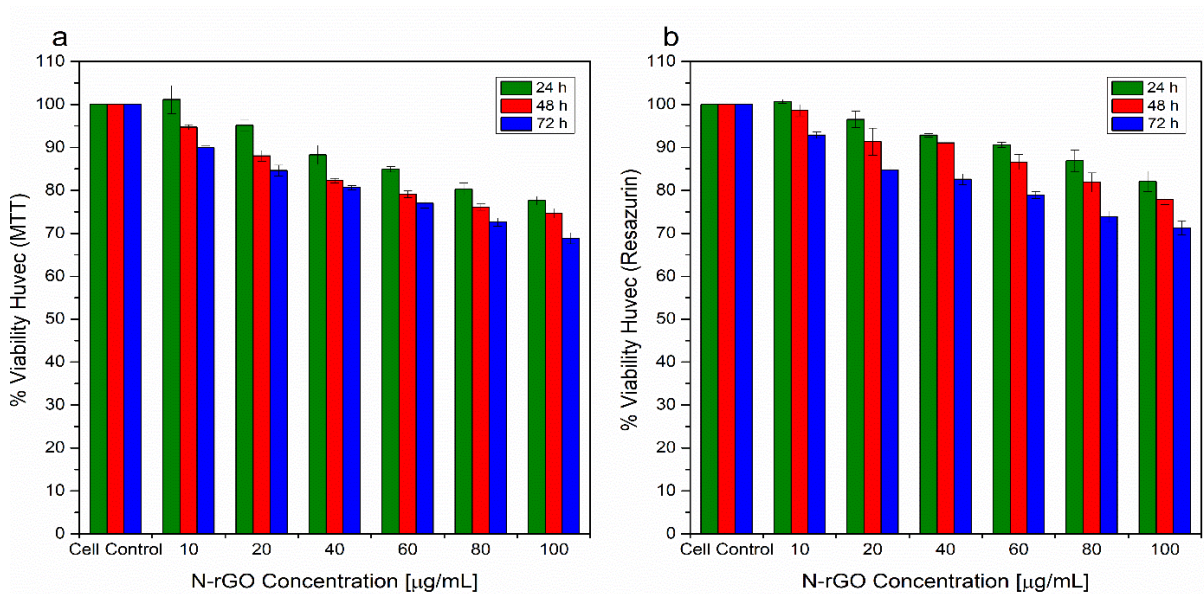

*Figure 13. % of viability a) Huvec with MTT assay, b) Huvec with resazurin assay.*

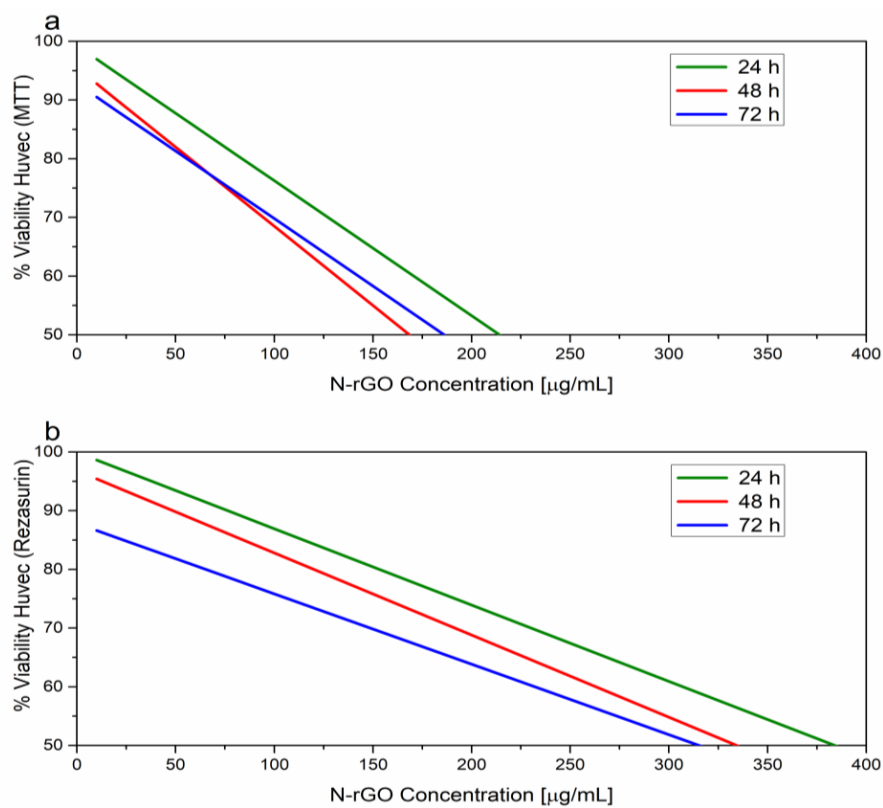

*Figure 14. IC50 a) HUVEC with MTT assay, b) PBMC with MTT assay, c) HUVEC with resazurin assay.*

**Conclusions**

An electrostatic repulsion mechanism for low density lipoproteins using reduced graphene oxide was investigated first time with this study. According to the obtained results, reduced graphene oxide doped with 9.57% at nitrogen was achieved by a chemical exfoliation process that involves few reagents and a low temperature process for the reduction, SEM, TEM and AFM characterizations confirmed few layers of reduced graphene oxide, XPS confirmed the presence of three types of nitrogen-carbon bonding: pyridinic N, pyrrole N and quaternary N, the predominant type was pyrrole N, ideal for the lipo-repellent application because it is the N bonding type that increases the most the negative electric charge of N-rGO surface, characteristic necessary to generate the repulsion of the also negative electric charged LDL. The reduction process also achieved a good percentage of oxygen reduction and lattice restoration. The integrity of the N-rGO after contact with biological fluid, specifically with LDL was proved by FTIR showing no significant changes in the structure, and the repulsion behavior between N-rGO and LDL was ratified by the obtention of high contact angle measurements (95.7° to 130.4°) for different LDL oxidation levels. Also, the negative electric surface charged of N-rGO given by the reduction/doping process implemented was confirmed by EFM complementing the contact angle results and corroborating that the liporepellent behavior is given by the nitrogen introduction. Finally, the cytotoxicity was proved correct on Huvec cell line for N-rGO concentrations up to 100 μg/mL. Considering the presented results, nitrogen doped reduced graphene oxide can be a potential reinforcement material for a stent or stent covering with the purpose of improve atherosclerosis treatment.

## Acknowledgments

The financial support by CONACyT. We thank to Nayely Pineda Aguilar, Alejandro Arizpe Zapata, Oscar E. Vega Becerra, F. Enrique Longoria Rodriguez, Patricia Cerda Hurtado, Lilia Magdalena Bautista Carrillo and L. Gerardo Silva Vidaurri from CIMAV Monterrey for the technical support.

## Conflicts of Interest

The authors declare no conflict of interest.